\documentclass[11pt]{IEEEtran} 
\usepackage[utf8]{inputenc}
\usepackage[noadjust]{cite}

\usepackage{graphicx}
\usepackage{amsmath}
\usepackage{amssymb}
\usepackage{mathtools}

\usepackage{nicefrac}
\usepackage{placeins}
\usepackage[hidelinks]{hyperref}
\usepackage{blindtext}
\usepackage{tikz}
\usetikzlibrary{external}
\tikzsetexternalprefix{external_fig/}
\pgfdeclarelayer{background}
\pgfsetlayers{background, main}
\usepackage[normalem]{ulem}
\usepackage{pgfplots}
\usepgfplotslibrary{units}
\usetikzlibrary{arrows}
\usetikzlibrary{automata,positioning}
\usetikzlibrary{shapes.geometric} 
\usetikzlibrary{shadows.blur}
\usetikzlibrary{shapes.symbols}
\usetikzlibrary{calc,decorations.markings,math,arrows.meta}
\usetikzlibrary{backgrounds}
\usetikzlibrary{patterns}
\usetikzlibrary{patterns.meta}
\usepackage{multirow}
\definecolor{leman}{RGB}{0,167,159}
\definecolor{commblue}{RGB}{0,96,172}
\definecolor{vertdeau}{RGB}{194,221,176}
\definecolor{perle}{RGB}{202,199,199}
\definecolor{canard}{RGB}{0,116,128}
\definecolor{groseille}{RGB}{181,31,31}
\definecolor{montrose}{RGB}{243,152,105}
\definecolor{zinzolin}{RGB}{92,36,131}
\definecolor{carotte}{RGB}{236,102,8}
\definecolor{chartreuse}{RGB}{200,211,0}
\definecolor{robinEggBlue}{RGB}{10,195,195}
\definecolor{milanoRed}{RGB}{195,5,5}
\definecolor{ardoise}{RGB}{69,58,76}

\definecolor{mygreen}{RGB}{50, 173, 26}
\definecolor{mygray}{RGB}{30, 30, 30}

\newif\ifshowdeleted
\newif\ifshowmodified
\newif\iftrackedchanges

\trackedchangesfalse
\ifdefined\trackedchangesflag
  \trackedchangestrue
\fi

\iftrackedchanges
	\showdeletedfalse 
	\showmodifiedtrue
\else
	\showdeletedfalse
	\showmodifiedfalse
\fi

\ifshowmodified
	\newcommand{\modified}[1]{\textcolor{commblue}{#1}}
	\else
	\newcommand{\modified}[1]{#1}
\fi

\ifshowdeleted
	\newcommand{\removed}[1]{\textcolor{groseille}{\sout{#1}}}
\else
	\newcommand{\removed}[1]{}
\fi
\pgfplotsset{compat=1.18}

\newcommand{\IEEEcopyrightnotice}{%
    \IEEEpubid{%
        \parbox[b]{\columnwidth}{%
            \scriptsize
            \copyright~2026 IEEE. Personal use of this material is permitted.  Permission from IEEE must be obtained for all other uses, in any current or future media, including reprinting/republishing this material for advertising or promotional purposes, creating new collective works, for resale or redistribution to servers or lists, or reuse of any copyrighted component of this work in other works.
        }%
        \hspace{\columnsep}%
        \makebox[\columnwidth]{}%
    }%
}
\newcommand{\reducebiovspace}{\vskip -2.45\baselineskip plus -1fil}
\tikzexternaldisable
\begin{document}
\author{\IEEEauthorblockN{Joachim Tapparel\IEEEauthorrefmark{1}, Amavi Dossa\IEEEauthorrefmark{2}, El Mehdi Amhoud\IEEEauthorrefmark{2} and Andreas Burg\IEEEauthorrefmark{1}}\\
	\normalsize\IEEEauthorblockA{\IEEEauthorrefmark{1}Telecommunications Circuits Laboratory, \'Ecole Polytechnique Fédérale de Lausanne, Switzerland\\
	\IEEEauthorrefmark{2}College of Computing, Mohammed VI Polytechnic University, Morocco
	}
} 
\title{Centralized RAN for Future Low-Power\\ Wide-Area Networks{: A LoRa Case Study}}
\IEEEoverridecommandlockouts
\IEEEaftertitletext{\vspace{-2em}}
\IEEEcopyrightnotice
\maketitle
\IEEEpubidadjcol
\IEEEpubidadjcol

\begin{abstract}
In recent years, low-power wide-area network~\mbox{(LPWAN)} technologies have gained significant traction as a connectivity option for Internet of Things~(IoT) applications.
While these networks have been successful in providing long-range, low-power, and low-cost connectivity, they currently face scalability, reliability, and efficiency challenges that require immediate attention.
In this paper, we first identify important challenges for \mbox{LPWANs}.
We then advocate for the introduction of a centralized radio access network~(\mbox{C-RAN}) architecture tailored for \mbox{LPWANs} and present a proof-of-concept implementation and deployment of the proposed \mbox{C-RAN} for the widely popular long range~(LoRa) standard.
We also provide experimental results to demonstrate and quantify the increased sensitivity that can be obtained from joint processing of the baseband signals of multiple receivers, enabled by the proposed centralized architecture \modified{in quasi-static scenarios and drone-mounted transmitters}.
\end{abstract} 

\section{Introduction}
Connected Internet of Things~(IoT) devices have improved the efficiency of a wide range of applications.
From smart cities to precision agriculture, the access to numerous connected sensors has brought new insights and has enabled unprecedented optimizations in many fields.
A new type of wireless connectivity, called low-power wide-area networks~(\mbox{LPWANs}), has been instrumental for providing access to the data gathered by many remotely-deployed and battery-powered sensors, with an estimation of more than $300$ million \mbox{LPWAN} devices shipped in 2024 alone.
Contrary to traditional cellular networks, \mbox{LPWANs} are designed to provide long-range, low-power, and low-cost connectivity for numerous devices.
Multiple technologies have been developed to provide such connectivity, with two main categories emerging: cellular-based and non-cellular-based.
Cellular \mbox{IoT} technologies, such as Narrowband IoT~\mbox{(NB-IoT)} and LTE-M, leverage existing cellular infrastructure and provide a high reliability and quality of service at the cost of a lower energy efficiency and higher complexity, also for the \mbox{IoT} devices.
To further improve the battery life and to reduce the operation costs, alternative technologies operate in the unlicensed industrial, scientific, and medical~(ISM) frequency bands with significantly simpler protocols and modulations~\cite{wu2023large}.
\par
In the following, we focus on the latter, non-cellular-based \mbox{LPWAN} technologies as their reliability faces many challenges due to their low-cost simple protocols and lack of an overseeing entity while sharing limited time and frequency resources.
These \mbox{LPWAN} technologies, such as the long range~(LoRa) standard, Sigfox, or Mioty allow users to deploy their own gateways to establish private networks or extend coverage of existing infrastructure.
For each network, user-deployed gateways are connected to a central server in a star topology in which the gateway is an independent entity responsible for receiving any message within its range.
While this type of deployment performs well for a moderate number of devices, the rapid expansion and proliferation of \mbox{IoT} networks and the need for long-range communication are challenging the capacity, the reliability, and the energy efficiency of the \mbox{LPWANs} end devices~\cite{jouhari2023survey}.
\par
\begin{figure*}[t]
	\centering
	\includegraphics[width=1\linewidth]{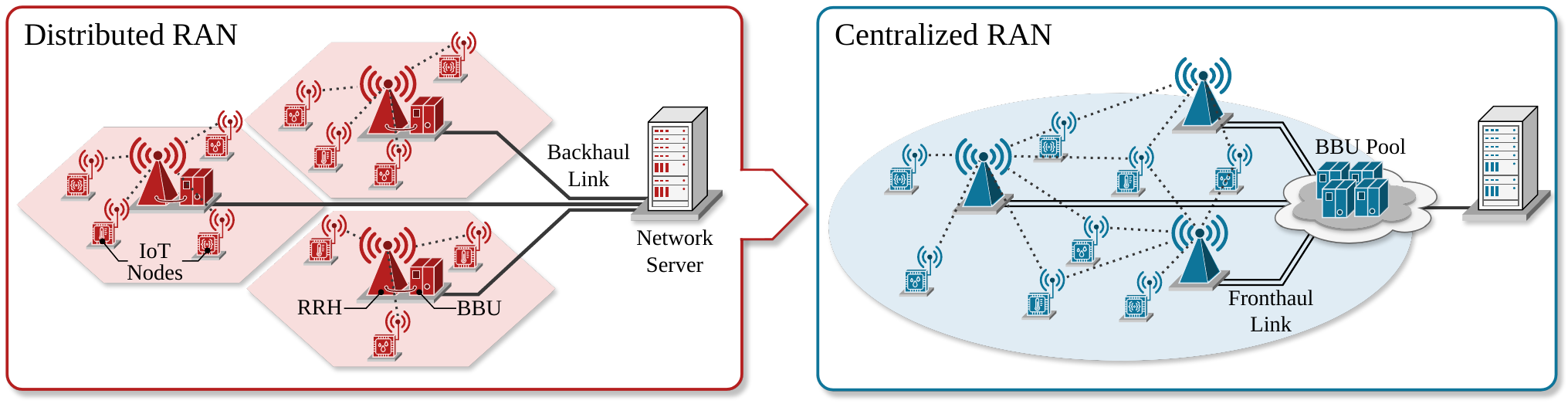}\\
	\vspace*{0pt}
	\caption{Illustration of a distributed and a centralized radio access network for LPWANs}
	\label{fig:dranCran}
\end{figure*}

   
In this paper, we first discuss critical challenges that \mbox{LPWANs} have to address to remain a leading and scalable technology for \mbox{IoT}.
We then motivate the use of a centralized radio access network~(C-RAN) architecture, illustrated in \figurename~\ref{fig:dranCran}, to overcome many of these issues.
The main focus is on improving the uplink transmissions of the IoT nodes, as they represent the vast majority of the traffic in \mbox{LPWANs}.
\modified{Building on the software stack in~\cite{tapparel2024cran}, we study the system-level benefits and deployment considerations of centralized processing and quantify the resulting sensitivity and mobility-robustness gains through campus-scale and drone-based LoRa measurements.}
Both the software stack and the dataset gathered through the measurement campaign on the EPFL campus in Lausanne, Switzerland, are available online~\cite{lora_cran_tcl_webpage}.

\section{{Critical Challenges Faced by LPWANs}}\label{sec:challenges}
The number of connected \mbox{IoT} devices has been estimated to grow from $10$ to $41$ billion between $2021$ and $2027$~\cite{elgazzar2022revisiting}.
This growth leads to numerous issues that \mbox{LPWANs} must address to remain a leading technology for \mbox{IoT}.
In the following, we outline the key technical challenges that arise both from the growing number of devices and from the diversification of use cases.

\subsection{Higher Sensitivity Requirement}
Coverage is a key performance factor for \mbox{LPWANs}, as sensors must operate across diverse environments ranging from dense urban areas to remote rural or indoor locations.
The physical layer of all relevant \mbox{LPWAN} technologies already achieves detections at received power levels that are much lower than those of conventional networks by using low and adaptive data rates.
However, sensitivity, defined as the minimal power level required for successful signal reception, still needs to improve to further increase coverage and network capacity.
Any improvement of the sensitivity benefits the network in multiple ways.
First, the coverage of the network can be significantly extended, as the signal can be detected at a lower power level~{\cite{ikpehai2019low}}.
Second, the network capacity is increased, as transmitters can use higher data rate modes which reduce both the time-on-air and the probability of collisions.
Finally, the energy consumption of the nodes is reduced, as the radio can use a lower transmission power or transmit over a shorter duration.

\subsection{Interference Mitigation}
The continued densification of \mbox{IoT} devices creates networks with a rapidly increasing amount of interference~\cite{georgiou2017low,tapparel2021enhancing,jouhari2023survey}.
Devices that use the ISM frequency bands are only constrained by very high-level regional regulations such as maximum transmission power and duty-cycle.
Because there is no central entity coordinating the frequency band or the devices of a given technology, network densification leads to two types of interference: inter-technology interference and intra-technology interference.
The first type of interference can mostly be considered as a global increase in white Gaussian noise at the receiver due to the spread spectrum or ultra narrowband modulations used by the most popular \mbox{LPWANs}.
Therefore, inter-technology interference can be combated with improvements in receiver (gateway) sensitivity.
The intra-technology interference is caused by devices of the same technology that rely on low overhead medium access control~(MAC) schemes.
The most popular non-cellular-based LPWAN protocols employ ALOHA-based scheduling, which does not impose any coordination or listen-before-talk overhead on the \mbox{IoT} nodes.
However, due to its simplicity and consequently low energy demands, ALOHA achieves only $18$\,\% of the throughput of an ideally scheduled channel, at an optimal network load.
While more complex detection methods, such as multi-user receivers, can increase the usable throughput to $48$\,\% of the channel capacity~\cite{tapparel2021enhancing}, the additional required processing can prove challenging for low-cost gateways. 
\subsection{Mobility Support}\label{sec:mobility}
When IoT sensors are deployed on mobile platforms such as drones, bicycles, or trucks, the wireless channel becomes time-varying. 
LPWAN transmissions are especially vulnerable because their low data rates lead to frame durations that can reach several seconds, while link adaptation cannot react within a frame. 
As a result, the frame duration can greatly exceed the channel coherence time, allowing multiple deep fades to occur during the reception of a single frame, as also observed during the measurement campaign described later in Sec.~\ref{sec:measurements}.
Even at a modest speed of $10$\,km/h, the coherence time at $868$\,MHz is on the order of $30$\,ms. 
Each receiver therefore requires a large received-power margin to decode the full frame on its own, which reduces energy efficiency \modified{of the IoT node}.
\subsection{Malicious Transmissions and Jamming Mitigation}
The utilization of unlicensed frequency bands opens opportunities for malicious transmitters to disrupt reliable transmissions while still complying with regulatory limits.
The receiver should be able to detect the presence of malicious users and mitigate their impact.
As the range of attacks is broad, from reactive jamming to energy drain, the receiver should have flexibility to identify the type of attack and establish suitable countermeasures.

\section{IoT C-RAN with Centralized Processing}
We propose to leverage the capabilities of a \mbox{C-RAN} architecture to address the challenges faced by \mbox{LPWANs}.
The opportunities of a centralized and virtualized processing of baseband signals~\cite{wang2017vran,chen2024evolution} can be leveraged to address some unique challenges of \mbox{LPWANs}, such as densification in pure ALOHA networks or fading mitigation within single long data frames that are sent without sophisticated modulation/coding schemes, scheduling, or fast rate adaptation.
While the \mbox{C-RAN} concept was originally designed for cellular networks, it must be adapted to \mbox{LPWAN} environments, which lack coordination between devices and scheduling capabilities.
However, the differences rather correspond to simplifications while still offering many benefits.
In the following, we describe the vision and a potential architecture for an \mbox{IoT} \mbox{C-RAN}\@.

\subsection{Technical Description}
\figurename~\ref{fig:dranCran} illustrates the architecture of an \mbox{IoT} \mbox{C-RAN}, where remote radio heads~(RRHs) only stream largely un-processed baseband samples to a central location for joint processing.
Rather than a set of distributed gateways, the network becomes a cell-free single-input multiple-output~\mbox{(CF-SIMO)} system.
The increased cooperation of the \mbox{RRHs} can address many challenges faced by \mbox{LPWANs} without any change to the \mbox{IoT} nodes.
\modified{While the IoT nodes remain unaltered, the proposed \mbox{C-RAN} shifts part of the power, cost, and processing effort to the network infrastructure, which must provision the fronthaul connectivity and the centralized BBU resources.}
\par
Due to their similar target range and energy consumption, different \mbox{LPWAN} technologies (e.g., LoRa, Sigfox, or Mioty) present very similar requirements for an \mbox{IoT} \mbox{C-RAN} architecture.
\modified{The proposed architecture is therefore not specific to LoRa, even though LoRa is used as the case study in this article. 
At the architectural level, the fronthaul transport, buffering, resource orchestration, and virtualized BBU framework can be reused across LPWAN technologies.
The RF configuration and waveform-dependent signal-detection, synchronization, demodulation, decoding, and packet-processing modules must be adapted to the target technology.}
\modified{\\
The fronthaul link has to transmit complex baseband samples with sampling rates in the order of a few hundred kilohertz, depending on the LPWAN technology~\cite{tapparel2024cran}.
For example, continuously streaming a typical $125$\,kHz LoRa signal with $8$-bit I and Q components requires a raw IQ payload rate of $2$\,Mbit/s.
In practical deployments, this load can be significantly reduced through lower-resolution quantization, compression, or activity detection at the RRH, such that IQ samples are transmitted only when a signal of interest is detected.}
\par
\modified{
The main fronthaul latency constraint arises because LPWAN devices listen for downlink responses only during receive windows following an uplink transmission.
Existing LPWAN deployments already rely on centralized network servers to process uplinks and schedule the corresponding MAC-layer commands or acknowledgements, rather than generating these responses locally at the gateway.
Accordingly, LPWAN protocols provide a delay of at least one second between the uplink and the corresponding downlink receive window to accommodate network transport and server processing~\cite[Tab.~1]{tapparel2024cran}.
The proposed architecture operates within the same timing budget, while requiring a moderate fronthaul bandwidth of a few Mbit/s.
These requirements are compatible with widely available consumer broadband connections, allowing them to serve as fronthaul links for individual RRHs.
}

\subsection{Benefits}
We now discuss how an \mbox{IoT} \mbox{C-RAN} architecture helps to address the challenges identified in Section~\ref{sec:challenges} and supports the future development of \mbox{IoT} networks.
\par\textit{Increased Sensitivity:}
The \mbox{C-RAN} architecture enables centralized processing of IQ samples collected from multiple \mbox{RRHs}. Since each \mbox{RRH} receives the same transmitted signal corrupted by largely uncorrelated noise, combining these signals centrally allows the use of techniques that are commonly employed in multi-antenna systems to obtain a better combined SNR\@. 
As further discussed in Section~\ref{sec:measurements}, methods such as maximal-ratio combining~(MRC) or other forms of joint estimation with relaxed coherence constraints can exploit these independent signal paths to attenuate noise. 
This noise reduction leads to an improvement in receiver sensitivity, thereby extending network coverage or enabling lower transmission power for the end devices.

\par\textit{\modified{Collision and Interference Mitigation:}}
\modified{
A \mbox{C-RAN} architecture renders IQ streams from multiple geographically separated \mbox{RRHs} available at the \mbox{BBU}.
Joint processing of spatially distributed observations has been widely investigated as a mean to improve interference resilience in cell-free \mbox{MIMO} networks~\cite{ammar2022usercentric}.
Building on this principle, the proposed architecture creates opportunities for multi-user detection and collision-recovery algorithms that exploit spatial diversity, including interference-aware combining across \mbox{RRHs} and joint successive interference cancellation.
Centralized processing further allows computationally demanding algorithms to be employed without replicating the corresponding processing hardware at each low-cost \mbox{RRH}.}

\par
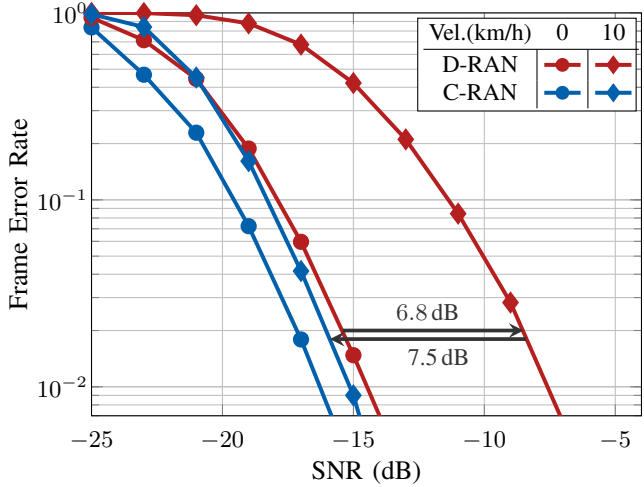
\begin{figure}
	\tikzsetnextfilename{diversity_gain_merged_single_match_drone}
	\tikzset{
  ModifiedBG/.style={
    show background rectangle,
    background rectangle/.style={fill=none, fill opacity=0.4}
  }
}
\begin{tikzpicture}[show background rectangle,
    background rectangle/.style={fill=none}]
\pgfplotsset{
compat=1.11,
}
\begin{axis}[%
width=\linewidth,
height=.78\linewidth,
name=main,
xmin=-25,
xmax=-4,
xlabel={SNR (dB)},
ymode=log,
ymin=7e-3,
ymax=1,
yminorticks=true,
ylabel={Frame Error Rate},
label style={font=\small},
xlabel style={yshift=2.9pt},
ylabel style={yshift=-4pt},
ticklabel style={font=\footnotesize},
xticklabel style={yshift=-2pt},
yticklabel style={xshift=2.5pt},
axis background/.style={fill=white},
xmajorgrids,
ymajorgrids,
yminorgrids,
legend columns=2, 
legend image post style={scale=0.8},
clip mode = individual,
]

\addplot [color=commblue,  line width=1.5pt, mark=*,mark size=2.3pt]
  table[row sep=crcr]{%
-25	0.836327345309381\\
-23	0.467065868263473\\
-21	0.22873\\
-19	0.07231\\
-17	0.01791\\
-15	0.00356\\
-13	0.00078\\
-11	0.0001\\
-9	2e-05\\
-7	2e-05\\
-5	0\\
-3	0\\
-1	0\\
};
\label{sf10_comb_0kmh}


\addplot [color=groseille, line width=1.5pt, mark=*,mark size=2.3pt]
  table[row sep=crcr]{%
-25	0.944111776447106\\
-23	0.714570858283433\\
-21	0.44365\\
-19	0.18815\\
-17	0.05964\\
-15	0.01476\\
-13	0.00331\\
-11	0.00055\\
-9	0.0001\\
-7	3e-05\\
-5	1e-05\\
-3	0\\
-1	0\\
};
\label{sf10_best_0kmh}

\addplot [color=groseille,, line width=1.5pt, mark=diamond*, mark size=3pt, mark options={solid}]
  table[row sep=crcr]{%
-25	1\\
-23	0.998003992015968\\
-21	0.972055888223553\\
-19	0.87933\\
-17	0.67684\\
-15	0.42152\\
-13	0.21067\\
-11	0.08439\\
-9	0.02822\\
-7	0.00649\\
-5	0.00124\\
-3	0.00016\\
-1	0\\
};
\label{sf10_best_10kmh}

\addplot [color=commblue,, line width=1.5pt, mark=diamond*, mark size=3pt, mark options={solid}]
  table[row sep=crcr]{%
-25	0.984031936127745\\
-23	0.840319361277445\\
-21	0.451097804391218\\
-19	0.1612\\
-17	0.0416\\
-15	0.009\\
-13	0.001\\
-11	0\\
-9	0\\
-7	0\\
-5	0\\
-3	0\\
-1	0\\
};
\label{sf10_comb_10kmh}

\addplot [color=commblue,line width=1.5pt, mark size=1.2pt, mark=none, mark options={solid,fill}]
  table[row sep=crcr]{-10 0.0001\\}; 
\label{cran}

\addplot [color=groseille,line width=1.5pt, mark size=1.2pt, mark=none, mark options={solid,fill}]
  table[row sep=crcr]{-10 0.0001\\}; 
\label{dran}

\draw[draw=black!80,->,>=stealth, line width=1.4pt] (-15.4,0.02) -- (-15.4+6.9,0.02) node[midway ,above,fill=white,fill opacity=0.8,yshift=2pt,xshift=-2pt,inner sep=1pt] {\footnotesize $6.8$\,dB};

\draw[draw=black!80,<-,>=stealth, line width=1.4pt,xshift=0pt] (axis cs:-15.9,0.018) -- (axis cs:-15.9+7.5,0.018) node[midway ,below,fill=white,fill opacity=0.8,yshift=-2pt,inner sep=1pt,xshift=4pt] {\footnotesize 7.5\,dB};

\end{axis}

        \node [draw,fill=white,inner sep=2pt,outer sep=2pt, at={(main.north east)},yshift=0pt,xshift=0pt,anchor=north east,inner sep=1pt,inner xsep=-1pt,text width = {}]{
      \footnotesize
      \setlength{\tabcolsep}{2pt}
      \begin{tabular}{c|c|c}
        \multicolumn{1}{c}{Vel.(km/h)}&\multicolumn{1}{c}{0}& 10\\
        \hline
        \raisebox{1pt}{\strut}D-RAN &\ref{sf10_best_0kmh} &\ref{sf10_best_10kmh} \\
                 C-RAN  & \ref{sf10_comb_0kmh} &\ref{sf10_comb_10kmh} \\
          \end{tabular}
          };

\end{tikzpicture}%
	\caption{FER of $30$-symbol LoRa frames using a spreading factor 10, code rate $\nicefrac{4}{5}$, and $250$\,kHz bandwidth with different mobility}
	\label{fig:mobility_cran_dran}
\end{figure}

\textit{Increased Diversity:}
As introduced in Section~\ref{sec:mobility}, the long duration of \mbox{LPWAN} frames increases the probability of outage during the reception of a frame.   
The multiple receiver locations provide uncorrelated signal paths that can be leveraged to mitigate the effect of channel fading.
To illustrate the available gains, \figurename~\ref{fig:mobility_cran_dran} presents the SNR gains obtained under low (walking-speed) mobility when combining the signals received by multiple \mbox{RRHs}.
The results are obtained from Monte Carlo simulations of LoRa transmissions using a Rayleigh fading channel model and four \mbox{RRHs} with the same average received power, but independent fading channels.
Four RRHs are chosen as they represent a good trade-off between the performance gain, which yields diminishing returns for each additional \mbox{RRH}, and the complexity of the system.
By comparing the static\,\eqref{sf10_best_0kmh} and the mobile\,\eqref{sf10_best_10kmh} cases, we observe a loss of $6.8$\,dB SNR with D-RAN even at $10$\,km/h.
In contrast, this loss is recovered when the signals from four RRHs are processed jointly in C-RAN compared to D-RAN\,\eqref{sf10_comb_10kmh}.
We note that coherent combining poses a number of challenges related to phase noise, jitter, or synchronization, and therefore, we only consider here the example of a non-coherent combining of log-likelihood ratios to avoid the need for inter-RRH phase coherence.
\par\textit{Jamming Mitigation:}
A commonly adopted approach to counter jamming in \mbox{LPWANs} is gateway densification.
By deploying multiple gateways in a specific area, the end-device signal is more likely to overpower the jamming signal at one of the gateways. 
\modified{
However, this approach becomes less effective in the presence of strong jamming signals or multiple jammers located near different gateways.
Different forms of diversity have been investigated in the context of jamming, particularly in multi-antenna systems, where independently received observations can be exploited for jammer detection and suppression~\cite{pirayesh2022jamming}.
The benefits of distributed reception have also been demonstrated for LoRa-based \mbox{C-RAN}, where centralized combining across multiple gateways improves jamming mitigation compared with distributed gateway processing~\cite{dossa2026cran}.
}

\section{C-RAN Architecture and Results}
In this section, we introduce the core elements of the proposed \mbox{IoT} \mbox{C-RAN} architecture for \mbox{LPWANs} and present the results of a measurement campaign conducted on the EPFL campus in Lausanne, Switzerland. 
\subsection{Architecture for LPWANs}
\begin{figure}[t!]
	\centering
	\includegraphics[width=.94\linewidth]{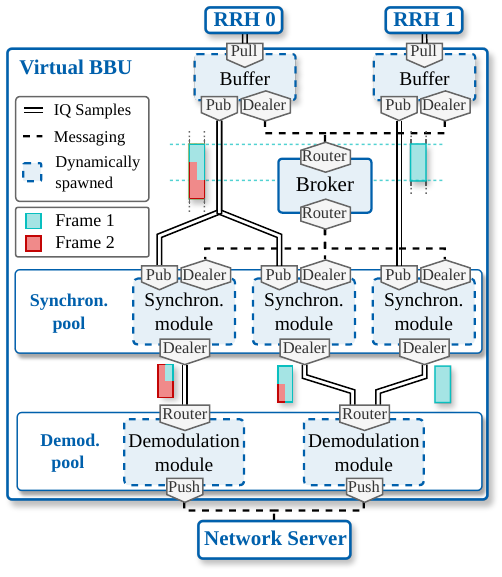}\\
	\vspace*{-5pt}
	\caption{Architecture of the \mbox{IoT} C-RAN with detailed BBU processing blocks and data flow for two example frames}
	\label{fig:cran_arch}
\end{figure}

\figurename~\ref{fig:cran_arch} illustrates our \mbox{IoT} \mbox{C-RAN} architecture for \mbox{LPWAN}\@.
The RAN is composed of three main parts: multiple \mbox{RRHs}, a network server, and a virtual BBU\@.
The \mbox{RRHs} perform a simple signal presence detection and only transmit the IQ samples that contain relevant samples, thereby avoiding unnecessary data usage.
Furthermore, this preliminary signal estimation at the \mbox{RRH} allows the processing load of the BBU to scale linearly with the number of detected signals rather than with the number of RRHs in the network.
The operations of the BBU are coordinated by a central broker, which handles the dynamic deployment of the processing blocks and associated buffers.
Additionally, the broker acts as a load balancer, distributing the processing workload of buffered IQ samples across available processing instances.
To prevent a bandwidth bottleneck, only control messages are routed through the broker and IQ samples are directly transmitted between the dedicated processing instances.
Each \mbox{RRH} connected to the BBU is assigned a dedicated buffer instance which stores the IQ samples that are received from the \mbox{RRH} until the IQ samples from different \mbox{RRHs} are available and the necessary resources are ready for processing.
We implement two types of processing entities: synchronization modules and demodulation modules.
The synchronization modules estimate and correct hardware impairments, such as carrier frequency offset and sampling frequency offset. While standard methods can be effective, techniques that are designed for the specific modulation of the \mbox{LPWAN} provide increased precision.
The demodulation modules can then further process the extracted frames to demodulate each symbol and decode the message content.
This functional separation enables high flexibility. 
For example, multiple synchronization modules can feed into a single demodulation instance to enable joint processing of concurrent IQ streams.
\modified{The radio-frequency synchronization required by these modules is performed independently for each received stream and does not imply tight time or phase synchronization among RRHs.
In the combining approach considered in the proposed architecture, each RRH signal is synchronized independently before the BBU combines post-synchronization information, rather than coherently adding phase-aligned IQ samples. 
Coarse time synchronization or timestamping is nevertheless useful to associate IQ segments received by different RRHs with the same transmission.}
\par
The virtual BBU shown in \figurename~\ref{fig:cran_arch} is populated with the necessary elements to process and combine the signals of two \mbox{RRHs}.
We illustrate the processing of two colliding frames received by \mbox{RRH} $0$ and the combining of one frame received by both \mbox{RRHs}.
Synchronization modules subscribe to buffer streams via a publish-subscribe protocol, which supports parallel processing of multiple transmissions within shared IQ data, such as the overlapping frames $1$ and $2$.
This protocol allows streaming a single IQ sample stream per frequency channel, regardless of the number of concurrent transmissions.
Communication between synchronization and demodulation modules is managed via dealer-router sockets, to allow dynamic connections between the different processing units in each pool.
In the illustrated example, two synchronization modules forward the synchronized samples of frame\;$1$ to a common demodulation instance for combining the signals received by the two \mbox{RRHs}.
Finally, the demodulated payloads are sent to a network server, identical to the one used in current D-RAN architectures. 
Existing MAC-layer deployments, such as LoRaWAN, can remain unchanged, as the BBU forwards uplink messages like a standard gateway.
For downlinks, the BBU forwards messages to the RRH with the most favorable channel conditions to the target device.
\modified{Details of the LPWAN C-RAN prototype, including its software framework, message passing, and processing pipeline, are available in~\cite{tapparel2024cran}.}
\par\textit{Node Access Capacity:}
\modified{Based on the measured CPU load reported in~\cite{tapparel2024cran},}
a single CPU core can process up to $6$ LoRa frames per second when combining the three strongest RRHs.
We use this measured throughput as a processing capacity to estimate how many nodes a single CPU core can support.
For instance, we assume a per-device airtime of $30$\,s/day and a Poisson arrival process. 
This airtime limit is commonly enforced by fair-use policies in public LoRaWANs and is consistent with the IoT application usage reported in~\cite{navarro2020survey}.
Under these assumptions, a single CPU core can support more than $550$ devices sending $590$ messages per day (SF7, 16B payloads), while keeping the overload probability below~$10$\%.
\tikzexternalenable
\begin{figure*}[t]
	\centering
		\includegraphics{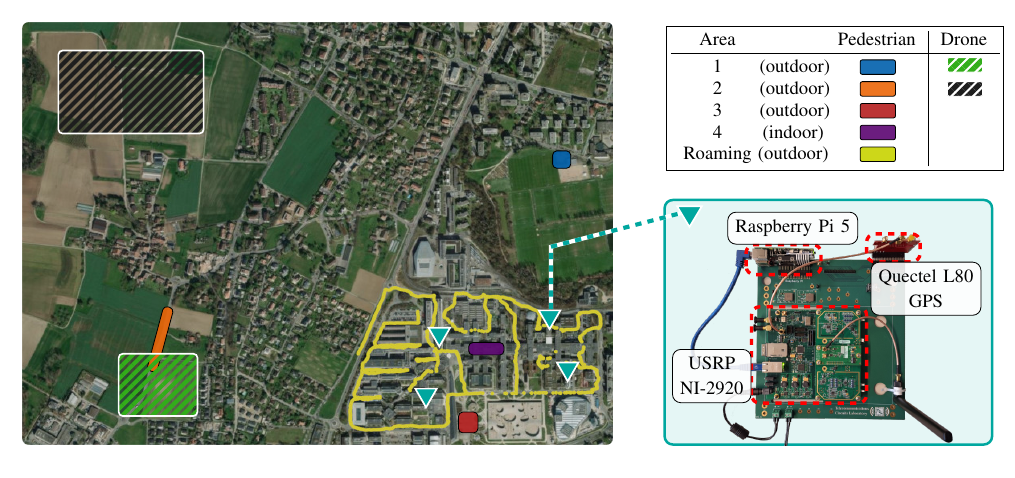}\\[-12pt]

 	\caption{Map of the measurement areas and RRH locations, together with RRH hardware details}
	\label{fig:measurementMap}
\end{figure*}
\tikzexternaldisable
\par\textit{\modified{Large-Scale BBU Processing Estimate:}}
\modified{Since the proposed \mbox{LPWAN} \mbox{C-RAN} performs packet-triggered processing, the BBU load scales mainly with the number of received frames.
Using the measured throughput of $6$ frames/s/core, we estimate the processing capacity required for a large-scale deployment handling on the order of $10^8$ messages per day, comparable to the public traffic reported by The Things Network.
This load requires about $200$ CPU cores without provisioning margin.
With a peak-to-average traffic factor of $2$ and a target CPU utilization of $80$\%, this estimate increases to about $480$ CPU cores.
Thus, the BBU processing requirement for a very large-scale \mbox{LPWAN} deployment is on the order of hundreds of CPU cores based on our current proof-of-concept implementation.
At the server level, this additional processing load would be distributed across several rack-mounted servers at regional network-server sites, depending on the hardware platform and redundancy margin.}
\subsection{{Measurements: LoRa Case Study}}\label{sec:measurements}
\begin{figure}
	\hspace{.03\linewidth}
		\includegraphics{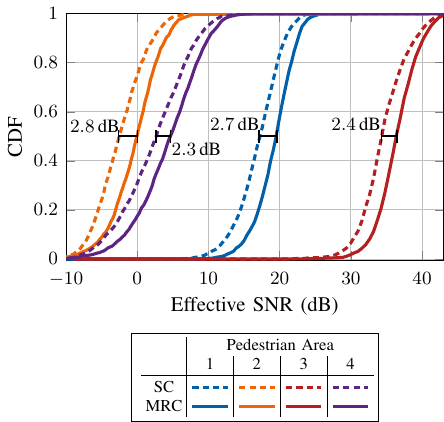}
	\caption{CDFs of the effective SNR under SC and ideal MRC for the quasi-static measurements
	}
	\label{fig:snr_cdf}	
\end{figure}

Our \mbox{LoRa} \mbox{C-RAN} prototype is implemented using GNU Radio, a popular open-source framework for building real-time software-defined radios, and ZeroMQ for messaging between the different RAN components.
The \mbox{RRHs} are built using a USRP NI-2920 controlled by a Raspberry Pi 5.
In addition, the \mbox{RRHs} are equipped with a Quectel L80 GPS module to provide a reference pulse per second~(PPS) signal for coarse time synchronization.
\modified{In the measurement campaign, this PPS signal mainly provides accurate timestamps for the reusable open-source dataset and simplifies the association of recordings from different RRHs.}\\
\figurename~\ref{fig:measurementMap} shows the hardware setup of the \mbox{RRHs} deployed for the experimental measurements as well as the location of the four \mbox{RRHs} installed on the EPFL campus.
The four \modified{pedestrian} measurement areas, also indicated on the map, are chosen to cover a variety of received signal-to-noise ratios~(SNRs) and different line-of-sight and non-line-of-sight conditions to the \mbox{RRHs}.
Areas $1$ and $2$ are located outside the campus with the first one having a clear path to the campus and the second one being obstructed behind a hill.
Areas $3$ and $4$ are on the campus, with the former being in an open outdoor area and the latter being indoor.
The LoRa transmitter used a spreading factor of $7$, a code rate of $4/5$, and payload size of 7 bytes.
These parameters are chosen to keep the frame duration short to limit the effect of fading within a frame, while also being representative of typical LoRa transmissions.
The frames were transmitted with a power of $14$\,dBm and a bandwidth of $125$\,kHz around an $862.5$\,MHz center frequency.
\modified{Two areas have been covered by a drone-mounted transmitter moving at a fixed $10$\,km/h to evaluate the gains for a mobile transmitter.
A larger spreading factor $10$ was used to transmit $19$\,B frames with a bandwidth of $250$\,kHz.}
The synchronization at the BBU is performed using the algorithm proposed in~\cite{xhonneux2022low}.
\paragraph{\modified{Available Combining Gains}}
We evaluate the sensitivity improvements enabled by the proposed \mbox{IoT} \mbox{C-RAN} architecture under quasi-static (low-mobility) channel conditions.
In a traditional \mbox{D-RAN} setup, selection combining~(SC) represents the practical upper bound of diversity gain only with no array-gain as each gateway processes signals independently. 
In contrast, the centralized nature of \mbox{C-RAN} allows for joint signal processing, enabling more advanced techniques such as MRC\@.  
\figurename~\ref{fig:snr_cdf} presents the cumulative distribution function~(CDF) of the received SNRs under both SC and ideal MRC across the four pedestrian measurement areas, showing the additional processing gains that are achievable through centralized processing.
The received signal power varies significantly between the different areas as the same transmission power was used for all the measurements.
However, independent of the area, the MRC provides a significant gain in the received effective SNR of $2.3$\,dB to $2.8$\,dB compared to SC\@.
An improvement of the post-combining SNR directly translates into an increase in sensitivity~\cite{jouhari2023survey}. 
While ideal combining of four equal-power signals with uncorrelated noise would yield at most a $6$\,dB gain, the observed gain is closer to that of combining two equal-power signals.
The obtained gains highlight that in a practical deployment, the few strongest \mbox{RRHs} contribute the most to the combining gain.
\tikzexternalenable
\tikzexternaldisable
\paragraph{\modified{Distribution of the Sensitivity Gains}}
To characterize the spatial distribution of the combining gain, we measured the reception of $10'000$ frames while slowly moving the transmitter within the EPFL campus, \modified{following the roaming path indicated in \figurename~\ref{fig:measurementMap}}.
\modified{For each received frame, we compare the effective SNR obtained with MRC to the one obtained with SC.
The measurements show that the combining gain is not limited to isolated locations, but is observed over a significant fraction of the roaming path.
In particular, the devices located along $60$\% of the roaming path could reduce their transmission energy by more than $20$\%, while still reaching the original, non-combined SNR level.
Among those devices, one third would benefit from more than a $33$\% reduction without any modifications to the devices themselves.
As the wireless transmission is one of the main contributors to the energy consumption of many \mbox{IoT} nodes, this reduction can directly improve the battery life of the devices.}

\begin{figure}
	\includegraphics{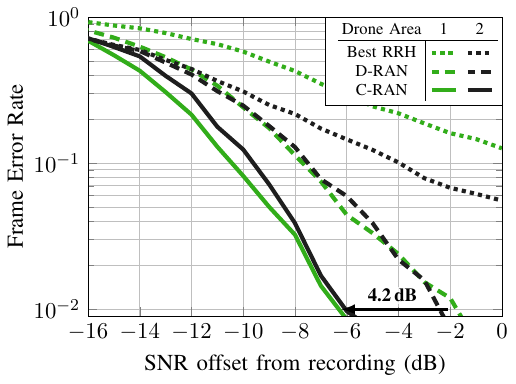}
	\caption{FER of centralized non-coherent combining compared with frame-level SC and the best independent RRH. The results are obtained for a transmitter velocity of $10$\,km/h, SF$10$, CR $\nicefrac{4}{5}$, and a $19$-byte payload}
	\label{fig:fer_for_snroffset}
\end{figure}

\paragraph{\modified{Gains for a Mobile Transmitter}}
\modified{To evaluate the benefit of centralized joint processing under realistic mobility, we recorded $2'000$ LoRa frames transmitted from a drone moving at a target speed of $10$\,km/h, received by $3.4$ RRHs on average.
The transmitter used SF10 and a $19$-byte payload.
\figurename~\ref{fig:fer_for_snroffset} shows the FER obtained after adding a set amount of independent white Gaussian noise to the recorded IQ samples before receiver processing.
This added noise provides a controlled post-recording SNR sweep while preserving the measured channel variation over time and relative received powers across the RRHs.
Compared with frame-level SC in a D-RAN architecture, the proposed C-RAN processing improves the reception robustness in both drone measurement areas.
In particular, at an FER of $1$\%, centralized non-coherent combining provides an SNR gain of up to 4.2 dB compared with D-RAN frame-level SC\@.
These measured mobile-transmitter results demonstrate that centralized multi-RRH processing improves robustness under time-varying mobile channels.}
\tikzexternalenable
\tikzexternaldisable
\section{Conclusion}
LPWANs face challenges related to densification, mobility, and coverage.
An \mbox{IoT} \mbox{C-RAN} architecture tailored to \mbox{LPWANs} enables the centralized processing of IQ samples, which improves the sensitivity through joint processing in both indoor and outdoor scenarios, mitigates channel fading, and increases the battery life of \mbox{IoT} devices.
\modified{Using measured per-RRH signal and noise powers, we estimate an ideal-MRC SNR gain of $2.3{-}2.8$\,dB over SC for the quasi-static measurements.
In the drone measurements, the implemented noncoherent receiver reduces the required SNR by about $4.2$\,dB compared with frame-level SC at an FER of $1\%$.}
Furthermore, the sensitivity increase applies to a large area, showing that a reduction in transmission energy of at least $20$\% is expected for nodes located in more than half of the evaluated locations.
These experiments highlight the potential of an \mbox{IoT} \mbox{C-RAN} architecture to address key \mbox{LPWAN} challenges and demonstrate the exciting possibilities for the next generation of \mbox{IoT} networks.
Future work could investigate MAC layer optimization strategies, multi-RRH interference mitigation, or fronthaul data compression to achieve further improvements in efficiency, coverage, and energy savings \modified{for the IoT nodes}.
Beyond the communication aspects, the centralized processing can also be beneficial for applications such as localization or sensing.

\section*{Acknowledgment}\noindent
The authors thank Dr.\,Wenqing Song for the valuable insights and constructive feedback on the manuscript.
This work was supported by the UM6P-EPFL Excellence in Africa Initiative (Junior Faculty Development program).

\bibliographystyle{IEEEtran}
\bibliography{IEEEabrv,paper.bib}

@STRING{IEEE_J_WCOML      = "{IEEE} Wireless Commun. Lett."}

@STRING{IEEE_J_IOT        = "{IEEE} Internet Things J."}

@STRING{IEEE_M_COM        = "{IEEE} Commun. Mag."}

@STRING{IEEE_O_CSTO       = "{IEEE} Commun. Surveys Tuts."}

@ARTICLE{wu2023large,
  author={Wu, Dixin and Bogdan, Alexandru S. and Liebeherr, Jörg},
  journal={IEEE Internet Things Mag.}, 
  title={Large-Scale Environmental Sensing of Remote Areas on a Budget}, 
  year={2023},
  volume={6},
  number={2},
  pages={130-136},
  doi={10.1109/IOTM.001.2200185}}

@ARTICLE{jouhari2023survey,
  author={Jouhari, Mohammed and Saeed, Nasir and Alouini, Mohamed-Slim and Amhoud, El Mehdi},
  journal=IEEE_O_CSTO, 
  title={A Survey on Scalable {LoRaWAN} for Massive {IoT}: Recent Advances, Potentials, and Challenges}, 
  year={2023},
  volume={25},
  number={3},
  pages={1841-1876},
}

@INPROCEEDINGS{tapparel2024cran,
  author={Tapparel, Joachim and Burg, Andreas},
  booktitle={IEEE Conf. NFV-SDN}, 
  title={Centralized {RAN} for {LPWAN}: Architecture and Proof-of-Concept Prototype Implementation}, 
  pages={1-6},
  year={2024},
  volume={},
  number={},
  doi={10.1109/NFV-SDN61811.2024.10807469}
  }

@Misc{lora_cran_tcl_webpage,
  author = {},
  title = {Centralized {RAN} Measurements of {LoRa}},
  publisher = {},
  journal = {},
  url = {https://www.epfl.ch/labs/tcl/resources-and-sw/lora-frames-dataset/},
  note ={{Accessed}: Aug. 24, 2026}
}

@ARTICLE{elgazzar2022revisiting,
author={Elgazzar, Khalid  and others },
title={Revisiting the {Internet of Things}: New trends, opportunities and grand challenges},
journal={Front. Internet Things},
volume={1},
year={2022},
doi={10.3389/friot.2022.1073780},
}

@ARTICLE{ikpehai2019low,
  author={Ikpehai, Augustine and others},
  journal=IEEE_J_IOT, 
  title={Low-Power Wide Area Network Technologies for Internet-of-Things: A Comparative Review}, 
  year={2019},
  volume={6},
  number={2},
  pages={2225-2240}
}

@article{georgiou2017low,
  title={Low power wide area network analysis: Can {LoRa} scale?},
  author={Georgiou, Orestis and Raza, Usman},
  journal=IEEE_J_WCOML,
  volume={6},
  number={2},
  pages={162--165},
  year={2017},
  publisher={IEEE}
}

@ARTICLE{tapparel2021enhancing,
  author={Tapparel, Joachim and Xhonneux, Mathieu and Bol, David and Louveaux, Jérôme and Burg, Andreas},
  journal={IEEE Open J. Commun. Soc.}, 
  title={Enhancing the Reliability of Dense {LoRaWAN} Networks With Multi-User Receivers}, 
  year={2021},
  volume={2},
  number={},
  pages={2725-2738},
  doi={10.1109/OJCOMS.2021.3134091}
}

@ARTICLE{wang2017vran,
  author={Wang, Xinbo and others},
  journal=IEEE_M_COM, 
  title={Virtualized Cloud Radio Access Network for {5G} Transport}, 
  year={2017},
  volume={55},
  number={9},
  pages={202-209},
}

@ARTICLE{chen2024evolution,
 author={Chen, Jiacheng and Liang, Xiaohu and Xue, Jianzhe and Sun, Yu and Zhou, Haibo and Shen, Xuemin},
  journal=IEEE_O_CSTO, 
  title={Evolution of {RAN} Architectures Toward {6G}: Motivation, Development, and Enabling Technologies}, 
  year={2024},
  volume={26},
  number={3},
  pages={1950-1988},
  doi={10.1109/COMST.2024.3388511}}

@ARTICLE{ammar2022usercentric,
  author={Ammar, Hussein A. and Adve, Raviraj and Shahbazpanahi, Shahram and Boudreau, Gary and Srinivas, Kothapalli Venkata},
  journal=IEEE_O_CSTO, 
  title={User-Centric Cell-Free Massive MIMO Networks: A Survey of Opportunities, Challenges and Solutions}, 
  year={2022},
  volume={24},
  number={1},
  pages={611-652},
  doi={10.1109/COMST.2021.3135119}}

@ARTICLE{pirayesh2022jamming,
  author={Pirayesh, Hossein and Zeng, Huacheng},
  journal=IEEE_O_CSTO, 
  title={Jamming Attacks and Anti-Jamming Strategies in Wireless Networks: A Comprehensive Survey}, 
  year={2022},
  volume={24},
  number={2},
  pages={767-809},
  doi={10.1109/COMST.2022.3159185}}

@ARTICLE{dossa2026cran,
  author={Dossa, Amavi and Tapparel, Joachim and Burg, Andreas and Amhoud, El Mehdi},
  journal={IEEE Open J. Commun. Soc.}, 
  title={A {C-RAN}-Based Mitigation of Reactive Jamming Attacks in {LoRa} Networks with Jamming Cancellation}, 
  year={2026},
  volume={7},
  number={},
  pages={8731-8747},
  doi={10.1109/OJCOMS.2026.3715389}}

@ARTICLE{navarro2020survey,
  author={Navarro-Ortiz, Jorge and Romero-Diaz, Pablo and Sendra, Sandra and Ameigeiras, Pablo and Ramos-Munoz, Juan J. and Lopez-Soler, Juan M.},
  journal=IEEE_O_CSTO, 
  title={A Survey on {5G} Usage Scenarios and Traffic Models}, 
  year={2020},
  volume={22},
  number={2},
  pages={905-929},
  doi={10.1109/COMST.2020.2971781}
}

@ARTICLE{xhonneux2022low,
  author={Xhonneux, Mathieu and Afisiadis, Orion and Bol, David and Louveaux, Jérôme},
  journal=IEEE_J_IOT, 
  title={A Low-Complexity {LoRa} Synchronization Algorithm Robust to Sampling Time Offsets}, 
  year={2022},
  volume={9},
  number={5},
  pages={3756-3769},
}
\reducebiovspace
\begin{IEEEbiographynophoto}{Joachim Tapparel} (Member, IEEE) received the B.Sc.\ and M.Sc.\ degrees in electrical engineering from the École Polytechnique Fédérale de Lausanne, Lausanne, Switzerland, in 2018 and 2021, respectively, where he is	currently pursuing the Ph.D. degree with the Telecommunications Circuits Laboratory.
His research interests include wireless communications for IoT systems, interference mitigation, and network architecture for wireless systems.
\end{IEEEbiographynophoto}
\reducebiovspace
\begin{IEEEbiographynophoto}{Amavi Dossa} (Student Member, IEEE) received his M.Sc.\ degree in electrical engineering from École Nationale Supérieure d'Électricité et de Mécanique, Casablanca, Morocco, in 2022. He is currently pursuing a Ph.D. at the Mohammed VI Polytechnic University, College of Computing, Benguerir, Morocco. His research interests include scalability and security of wireless IoT systems. 
\end{IEEEbiographynophoto}
\reducebiovspace
\begin{IEEEbiographynophoto}{El Mehdi Amhoud}(Member, IEEE) received the Ph.D. degree in computer and communication sciences from Télécom ParisTech, France. 
He is currently an Assistant Professor with Mohammed VI Polytechnic University, Morocco, and was previously a Postdoctoral Research Fellow with the King Abdullah University of Science and Technology, Saudi Arabia. 
He holds several U.S. patents. 
His research interests include modeling and analyzing the performance of new generations of communication networks and the Internet of Things.
He received two awards of excellence for his outstanding Ph.D. thesis from the Mines-Télécom Institute and the Marie Skłodowska-Curie Research Grant from the European Commission.
\end{IEEEbiographynophoto}
\reducebiovspace 
\begin{IEEEbiographynophoto}{Andreas Burg} (Senior Member, IEEE) 
received his Diploma and Ph.D. degrees from the Swiss Federal Institute of Technology (ETH) Z\"urich, Switzerland, in 2000 and 2006, respectively. 
During his doctorate, he worked one year at Bell Labs Wireless Research, NJ, USA. He was a Post-Doctoral Researcher at ETH Z\"urich (2006--2007) before co-founding Celestrius, an ETH spin-off in MIMO wireless communication. 
He returned to ETH Z\"urich as an SNF Assistant Professor in 2009 and moved to EPFL, Lausanne, in 2011, where he became a tenured Associate Professor in 2018. 
In 2021, he co-founded RAAAM Memory Technologies. 
He has served as TPC Co-Chair for VLSI-SoC 2012, ESSCIRC 2016, and SiPS 2017; General Chair of ISLPED 2019; and as an Editor for \textsc{IEEE Transactions on Circuits and Systems}, \textsc{IEEE Transactions on VLSI}, and \textsc{IEEE Transactions on Signal Processing}.
\end{IEEEbiographynophoto}
\reducebiovspace

\vfill
\end{document}